# Open-radiomics: A Collection of Standardized Datasets and a Technical Protocol for Reproducible Radiomics Machine Learning Pipelines

Khashayar Namdar[1,2,5,8], Matthias W. Wagner[1,2,3,4], Birgit B. Ertl-Wagner[1,2,3], Farzad Khalvati[1,2,3,5,6,7,8*]

[1]Division of Neuroradiology, Department of Diagnostic & Interventional Radiology, The Hospital for Sick Children (SickKids), Toronto, ON, Canada

[2]Neurosciences & Mental Health Research Program, SickKids Research Institute, Toronto, ON, Canada

[3]Department of Medical Imaging, University of Toronto, Toronto, ON, Canada

[4]Department of Diagnostic and Interventional Neuroradiology, University Hospital Augsburg, Germany

[5]Institute of Medical Science, University of Toronto, Toronto, ON, Canada

[6]Department of Computer Science, University of Toronto, Toronto, ON, Canada

[7]Department of Mechanical and Industrial Engineering, University of Toronto, Toronto, ON, Canada

[8]Vector Institute, Toronto, ON, Canada

# Abstract

**Background**
As an important branch of machine learning pipelines in medical imaging, radiomics faces two major challenges namely reproducibility and accessibility. In this work, we introduce open-radiomics, a set of radiomics datasets along with a comprehensive radiomics pipeline based on our proposed technical protocol to investigate the effects of radiomics feature extraction on the reproducibility of the results.

**Methods**
We curated large-scale radiomics datasets based on three open-source datasets; BraTS 2020 for high-grade glioma (HGG) versus low-grade glioma (LGG) classification and survival analysis, BraTS 2023 for O6-methylguanine-DNA methyltransferase (MGMT) classification, and non-small cell lung cancer (NSCLC) survival analysis from the Cancer Imaging Archive (TCIA). We used the BraTS 2020 open-source Magnetic Resonance Imaging (MRI) dataset to demonstrate how our proposed technical protocol could be utilized in radiomics-based studies. The cohort includes 369 adult patients with brain tumors (76 LGG, and 293 HGG). Using PyRadiomics library for LGG vs. HGG classification, we created 288 radiomics datasets; the combinations of 4 MRI sequences, 3 binWidths, 6 image normalization methods, and 4 tumor subregions. We used Random Forest classifiers, and for each radiomics dataset, we repeated the training-validation-test (60%/20%/20%) experiment with different data splits and model random states 100 times (28,800 test results) and calculated the Area Under the Receiver Operating Characteristic Curve (AUROC).

**Results**
Unlike binWidth and image normalization, tumor subregion and imaging sequence significantly affected performance of the models. T1 contrast-enhanced sequence and the union of Necrotic and the non-enhancing tumor core subregions resulted in the highest AUROCs (average test AUROC 0.951, 95% confidence interval of (0.949, 0.952)). Although several settings and data splits (28 out of 28800) yielded test AUROC of 1, they were irreproducible.

**Conclusions**
Our experiments demonstrate the sources of variability in radiomics pipelines (e.g., tumor subregion) can have a significant impact on the results, which may lead to superficial perfect performances that are irreproducible.

Keywords: radiomics, open-source, dataset, brain cancer, reproducibility

## List of abbreviations

AI: Artificial Intelligence
AT: Active Tumor
AUROC: Area Under the Receiver Operating Characteristic Curve
CI: Confidence Interval
CLT: Central Limit Theorem
CT: Computed Tomography
DICOM: Digital Imaging and Communications in Medicine
DWT: Discrete Wavelet Transform
ED: Edematous/Invaded Tissue
ET: Enhancing Tumor
GLSZM: Gray Level Size Zone Matrix
GTV: Gross Tumor Volume
HGG: High-Grade Glioma
IBSI: The Image Biomarker Standardization Initiative
LBP: Local Binary Pattern
LGG: Low-Grade Glioma
MGMT: O6-methylguanine-DNA methyltransferase
ML: Machine Learning
MRI: Magnetic Resonance Imaging
NCR: Necrotic Tumor
NET: Non-enhancing Tumor
NN: Neural Network
NIfTI: Neuroimaging Informatics Technology Initiative
NZV: Near-zero Variance
PCA: Principal Component Analysis
RF: Random Forest
ROC: Receiver Operating Characteristic
ROI: Region of Interest
RTSTRUCT: Radiotherapy Structure Set
SD: Standard Deviation
SVM: Support Vector Machines
TC: Tumor Core
TCIA: The Cancer Imaging Archive
VOI: Volume of Interest
WT: Whole Tumor

## Background

Artificial Intelligence (AI) has found its applications across different fields, and medical imaging is one of the high-potential areas where AI solutions are promising [1], [2]. Machine Learning (ML) is a subset of AI with tools for classification, regression, and decision-making, with many applications for medical imaging data [3]. ML algorithms in medical imaging fulfill tasks such as region of interest (ROI) segmentation and classification, and they are used as building blocks of AI-based pipelines for diagnosis, prognosis, and therapeutic assessments.

Deep Learning (DL) is a branch of ML where multiple-layer Neural Networks (NNs) are utilized at its core. Currently, AI-based segmentation is usually done using DL algorithms. However, for ML-based classification of medical imaging data, DL has a conventional competitor namely radiomics. The suffix "omics" refers to large-scale data derived to understand a biological perspective, such as genomics in Genetics [4]. Radiomics is a large set of manually defined features to study ROIs in Radiology images. From the Data Science point of view, radiomics provides a mapping, i.e., converting medical images into tabular data. Radiomics studies often start with image annotation, where 2D ROIs or 3D volumes of interest (VOIs) are segmented. Each radiomic feature is the result of applying a distinct and predefined equation to the ROI/VOI. Once the radiomic features are extracted, any ML classifier capable of handling tabular data, such as Random Forests (RF) [5], can be used to perform the classification task.

In comparison to DL, radiomics may offer a higher degree of explainability since its features are derived using transparent equations. However, there are multiple sources of variability impacting radiomics generalizability and reproducibility [6]. Radiomic features are sensitive to any form of change in ROIs/VOIs which leads to changes in the intensity values of image pixels/voxels within an ROI/VOI. Different vendors, imaging scanner settings, imaging protocols, contouring discrepancy (known as intra- and inter-reader variability), image normalization, and radiomics extraction settings may result in variation in radiomics results. Insufficient technical details, such as data split information, and lack of openness to the data and codes are other obstacles to having reproducible radiomics research. Depending on the source of the variability, addressing the issue may be infeasible in a given study. An example is tackling intra- and inter-reader variability in a fixed dataset of radiology images and ROI/VOI segmentations, without further information about the reader and/or not having access to annotation of other reader. Nevertheless, all these sources of variability create a demand for a proper statistical approach for measuring the randomness of the results and hence, the reliability of radiomics studies. Radiomics studies often lack systematic randomness measurements. Thus, in this research, the aim is to provide open-source radiomics datasets along with a baseline classification pipeline. We also propose a technical protocol for developing radiomics pipelines for reproducible radiomics-based classification models.

To demonstrate how the proposed technical protocol can improve reproducibility of radiomics-based studies, we use the BraTS 2020 [7] [8] [9], an open-source multimodal Magnetic Resonance Imaging (MRI) datasets for brain tumor segmentation. BraTS is primarily a segmentation dataset and hence, the majority of articles in the literature are dedicated to automated segmentation methods for brain tumors [7], [9], [10]. As for classification methods, Dequidt et al. [11] used BraTS 2018 to conduct a radiomics-based low-grade glioma (LGG) vs. high-grade glioma (HGG) classification task. Five expert radiologists labeled the tumors based on World Health Organization (WHO) standards, which enabled them to compare their model against another set of ground truths in addition to the BraTS labels. They extracted a limited set of radiomic features (51 features for each MRI sequence), used Support Vector Machines (SVM) [12] as the classifier with a 5-fold cross-validation for hyperparameter optimization, and achieved 84.1% accuracy with reference to the BraTS ground truths. Coupet et al. used BraTS 2020 as one of the datasets to train their models for healthy vs. glioma classification [13]. Polly et al. applied Otsu thresholding [14] to the images, used k-means for segmentation, Discrete Wavelet Transform (DWT) for feature extraction, and Principal Component Analysis (PCA) for dimensionality reduction, followed by SVM for a two-stage classification: abnormal vs. normal and then HGG vs. LGG. They achieve 99% accuracy on a small and balanced subset (50 HGG and 50 LGG) of BraTS 2017 and BraTS 2013 datasets, with a one-time data split approach. Integrating deep learning with radiomics has been shown to enhance the performance of radiomics pipelines [15][16]. However, this study prioritizes radiomics-only approaches due to their superior explainability.

In this paper, using BraTS 2020 dataset, we propose a comprehensive approach to investigating the effect of technical sources of variability for radiomics feature extraction including image normalization, the most impactful radiomics feature extraction hyperparameter (binWidth), imaging sequence, and tumor subregion on a radiomics-based tumor type classification pipeline. BraTS is an evolving collection of brain MRI datasets. Compared with BraTS 2020, BraTS 2021 included more patients and a O6-methylguanine-DNA methyltransferase (MGMT) classification challenge in addition to tumor segmentation. BraTS 2023 is an extension of BraTS 2021 for tumor segmentation. We extract radiomics features for BraTS 2023 and validate our proposed radiomics pipeline for reproducibility using the cohort from BraTS 2021 for MGMT classification. We also apply the proposed radiomics pipeline to computed tomography (CT) images of patients with non-small cell lung cancer (NSCLC) [17].

Contributions of this paper include a) providing large-scale radiomics datasets based on three open-source datasets; BraTS 2020 for HGG versus LGG classification and survival analysis, BraTS 2023 for MGMT classification [10] [7] [8], and NSCLC survival analysis from the Cancer Imaging Archive (TCIA) b) proposing open-radiomics technical research protocol, and c) providing a baseline for BraTS classification based on open-radiomics protocol as a technical validation.

## Methods

While there are annotated open-source medical imaging data, the extracted radiomics features are not usually available. Thus, each researcher must choose the appropriate tools/libraries and settings to extract the features. Consequently, radiomics studies are often conducted on small internal datasets with a selection of hyperparameters narrowed down for extracting the features. Open-radiomics (https://openradiomics.org) is our initiative for open-source large-scale radiomics datasets where we provide AI-ready tabular datasets along with baselines.

One of the sources of variability in radiomics research is the discrepancy between the feature extraction software and packages used across the studies. Multiple options are available to the research community for radiomics feature extraction [18]. Nonetheless, their back-ends may be different (e.g., different default settings or numerical precision), leading to irreproducibility of radiomics research. We will use PyRadiomics [19], which is widely used and supported by a large and established community. It should be highlighted that PyRadiomics-based packages, such as the Slicer Radiomics add-on module for 3D Slicer software

(https://www.slicer.org/) [20] may mask some features of PyRadiomics (e.g., Local Binary Pattern (LBP) features), resulting in suboptimal feature extraction.

## Datasets

### BraTS 2020

The dataset is a collection of multisequence MRIs, including T1-weighted (T1), gadolinium-based contrast agent enhanced T1-weighted (T1CE), T2-weighted (T2), and FLAIR sequences. The training cohort of BraTS 2020 is the data applicable to our research because its ground truth VOI segmentations are available. The cohort includes 369 adult patients with brain tumors, of which 76 cases are LGG, and 293 are HGG tumors. The images are all co-registered to the SRI24 atlas [21], skull stripped, resampled to 1 mm$^3$, and their size is unified to 155 × 240 × 240 voxels. We acknowledge that there is a discrepancy in the definition of LGG and HGG between BraTS and WHO [11]. In this paper, we follow the binary grading system (HGG vs. LGG) provided by BraTS dataset.

BraTS 2012-2016 included four tumor subregions, labeled 1-4. The necrotic (NCR), and the non-enhancing (NET) tumor core were labeled as 1 and 3, respectively. Label 2 corresponded to the peritumoral edematous/invaded tissue (ED), and active tumor (AT) was labeled as 4. Since BraTS 2017, NET and NCR are combined, and label 3 is removed from the annotations. Figure 1 depicts one slice of an image volume along with its corresponding segmentation mask. We analyze the whole tumor (all four subregions combined), AT, ED, and the union of NET and NCR (NETnNCR), in separate scenarios.

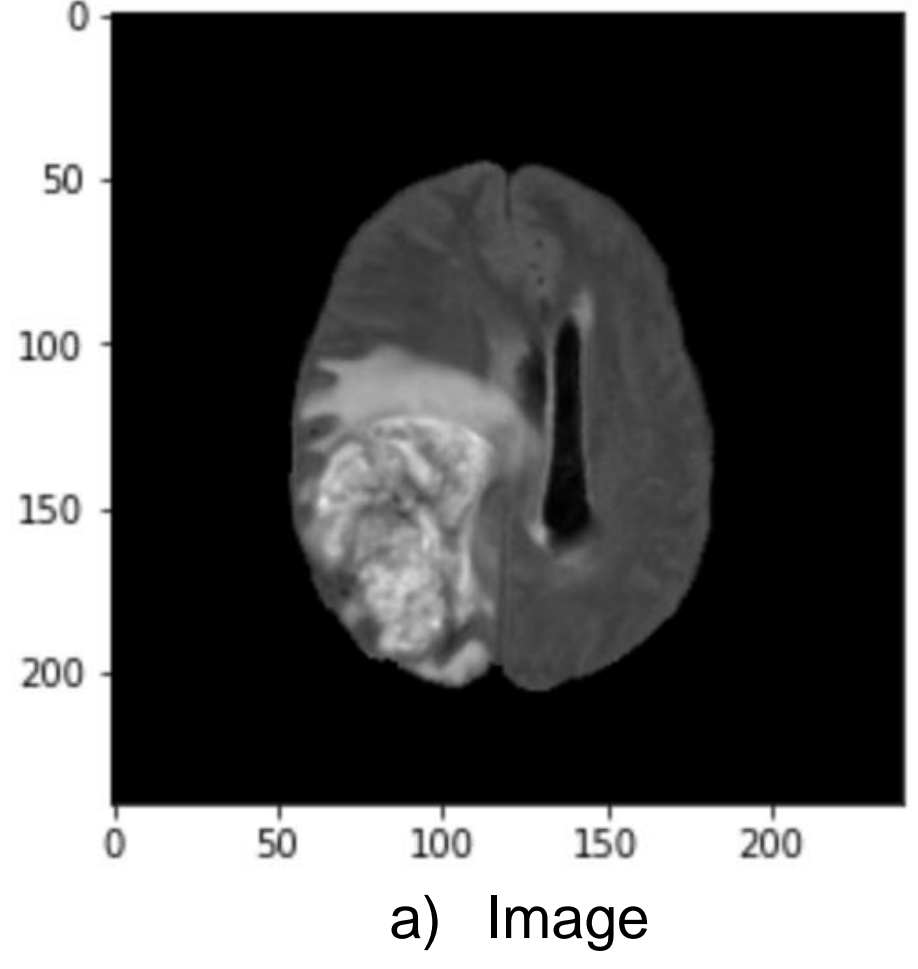

a) Image

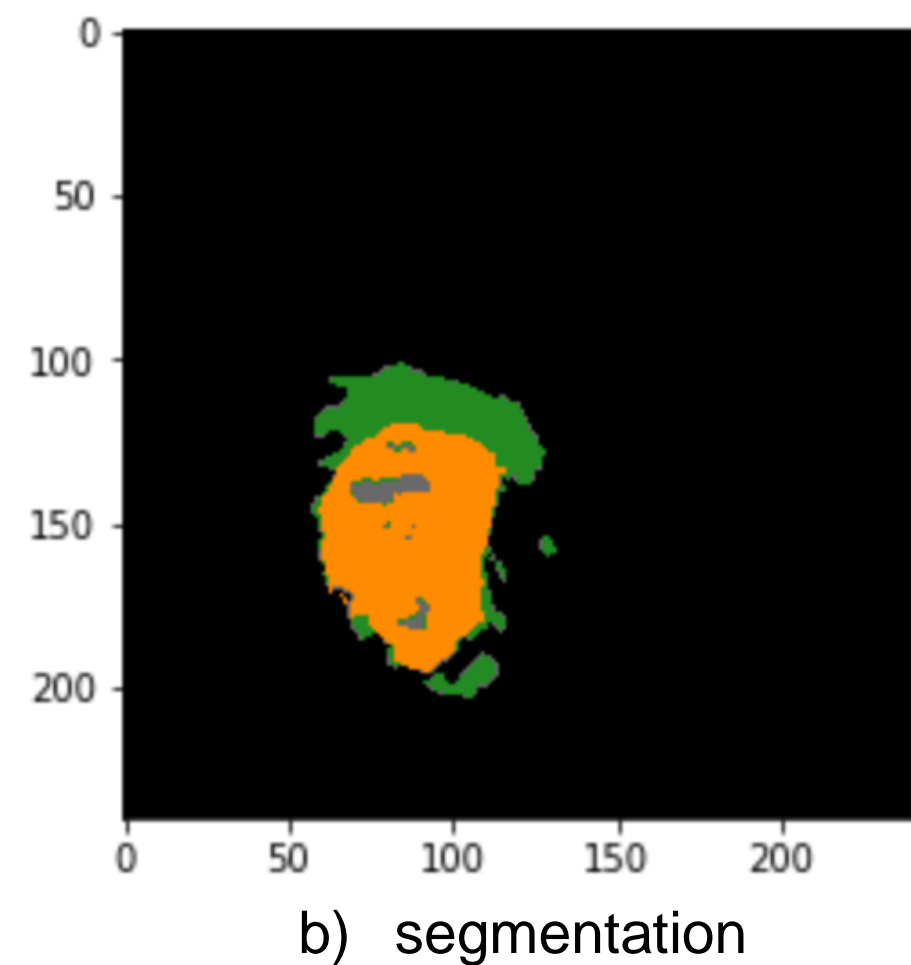

b) segmentation

**Figure 1:** An example BraTS 2020 image (the FLAIR sequence) and its corresponding segmentation mask. The orange area is AT, the green area is ED, and the gray parts are NETnNCR [7] [8] [9]

For 236 patients out of 369, survival information, gender, and extent of resection are available and are included in open-radiomics. Thus, open-radiomics BraTS 2020 supports both classification and survival analysis.

### BraTS 2023

BraTS 2023 is curated for segmentation of brain diffuse glioma patients and their sub-regions. Ground-truth segmentation masks are available for 1251 patients, with preprocessing procedure, imaging sequences, and tumor subregions similar to BraTS 2020. Through matching patient IDs, we included the MGMT classification ground-truth labels from BraTS 2021 in open-radiomics BraTS 2023. Out of 1251 patients, MGMT classification labels are available for 577, of which 301 have methylated MGMT (labeled as 1) and 276 have unmethylated MGMT (labeled 0). In terms of tumor subregions, BraTS 2023 includes enhancing tumor (ET), tumor core (TC), and the whole tumor (WT). TC entails the ET, and NCR. In open-radiomics BraTS 2023, we provide radiomics features extracted from WT, ED, ET, and NCR. As a result, open-radiomics BraTS 2023 includes 288 sets of radiomics for 1251 patients.

### TCIA NSCLC

This dataset comprises images from 422 patients diagnosed with NSCLC. For each patient, pretreatment CT scans are provided along with manually delineated 3D volumes of the gross tumor by a radiation oncologist, and associated clinical outcome data (i.e., survival). Open-radiomics NSCLC provides full radiomics features for the 422 patients in 18 sets (combinations of 3 binWidths and 6 image normalizations). The datasets include age, clinical stage, histology, gender, survival time, and dead/alive status event, and thus can be used for a range of ML tasks such as classification and survival analysis.

NSCLC dataset on TCIA provides digital imaging and communications in medicine (DICOM) format for the images and radiotherapy structure set (RTSTRUCT) for the segmentation masks. We converted the images and each segmentation mask to neuroimaging informatics technology initiative (NIfTI), which is widely supported by ML pipelines and libraries. Segmentation masks for gross tumor volume (GTV), left and right lungs, spinal cord, heart, and esophagus are available for the patients and can be used for training multi-class segmentation models (Figure 2). The preprocessed data is publicly available [22], [23].

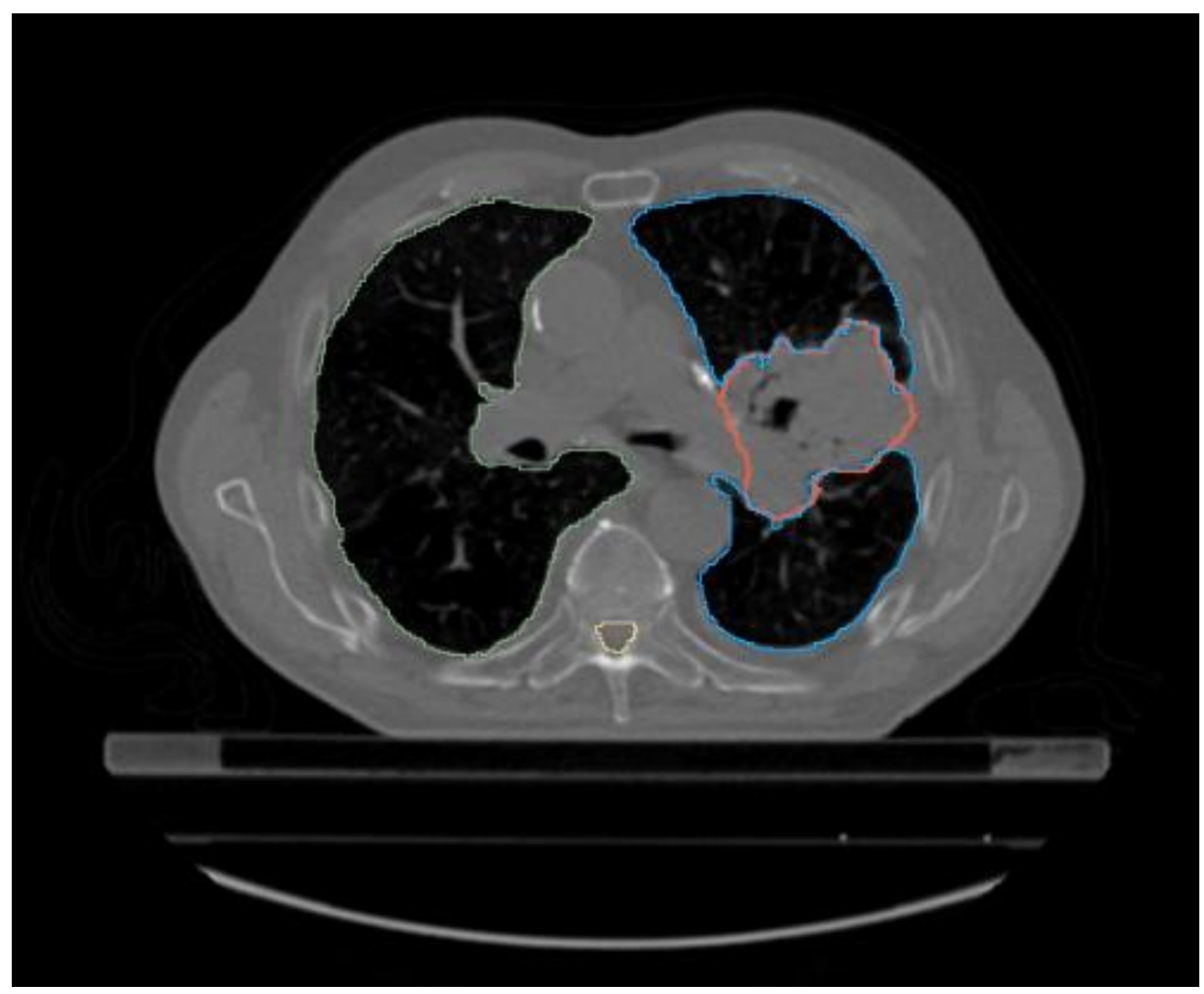

**Figure 2:** An example TCIA NSCLC CT image and its corresponding segmentation masks: left lung, right lung, spine, and GTV annotated with green, blue, yellow, and red, respectively [17].

## PyRadiomics Library

For radiomics feature extraction using PyRadiomics, there are technical details to be considered. Installation of the trimesh python package is essential (which can be done using pip install trimesh) to ensure all radiomic feature categories are extracted.

In general, radiomics studies can be 2D or 3D. A 2D analysis is used when the imaging method is 2D, such as X-Ray. However, when 3D images are accessible, 2D and 3D radiomics analyses are both possible. If the 2D approach is utilized with 3D images, the analysis is often done on the largest 2D cross-section of the ROI (e.g., tumor). In the case of 3D analyses, such as this study, enabling the full set feature extraction forces the PyRadiomics library to extract 2D LBP features [24] in addition to the 3D LBP features (see Appendix A). In this research, we included both 2D and 3D LBP features, and thus a full set of 1,710 features were extracted for each VOI.

PyRadiomics has multiple hyperparameters that affect the feature extraction procedure [25]. One of the most important is binWidth, which has been studied in the literature extensively [26][27][28]. binWidth determines the bin size which is needed to create histograms used for discretizing gray levels in the image, and thus affects all features except the shape features which are independent of pixel/voxel intensities. The default value of binWidth in PyRadiomics is 25. In this research, in addition to binWidth of 25, we examine 35 and 15 to see how they affect the results. All other hyperparameters are set to their default values[25].

## Image normalization

Image normalization plays an important role in ML pipelines, which can influence radiomics significantly. We implemented histogram equalization [29], z-score [30], gamma [31], and minmax [32] image normalization methods and incorporated them into our 3D analyses. As the coefficients of our gamma normalization, we explored 0.5 and 1.5, and the minmax normalization clipped voxel values of the image volumes between 0 and 1. All the normalization methods were applied to the image volumes, not the VOIs. We did not investigate VOI normalization (where only voxels in the VOI are used for normalization) in this research because it would increase the computational load. Nonetheless, it can be studied in the future.

## Classification pipeline

Once feature extraction is complete, the classification modules can be designed and implemented. With the default settings of PyRadiomics, each dataset has a group of diagnostics features (e.g., python version, simple itk version, etc.), which will not contain differentiative information, and thus we filtered them.

We propose a repetitive approach to measure randomness of our radiomics-based ML classification pipeline, which is illustrated in Figure 3. We repeat our evaluations on the test sets *N* times. In this research, *N* is set to 100. In the external for loop with *N* repetitions, in each iteration we randomly split our data into test (p%) and development (dev) sets. In this research, p was set to 20%. In the next step, a feature filtration algorithm is trained on the dev set and applied to the test set to randomly drop one of the features in pairs with a correlation coefficient above 0.95.

Another layer of feature selection is Near-zero Variance (NZV) filtration. We train an algorithm on the dev set and apply it to the test cohort to remove any feature with a variance lower than 0.05. The last step of feature manipulation is a MinMax scaler which learns the transformations from the dev set and applies them to the test.

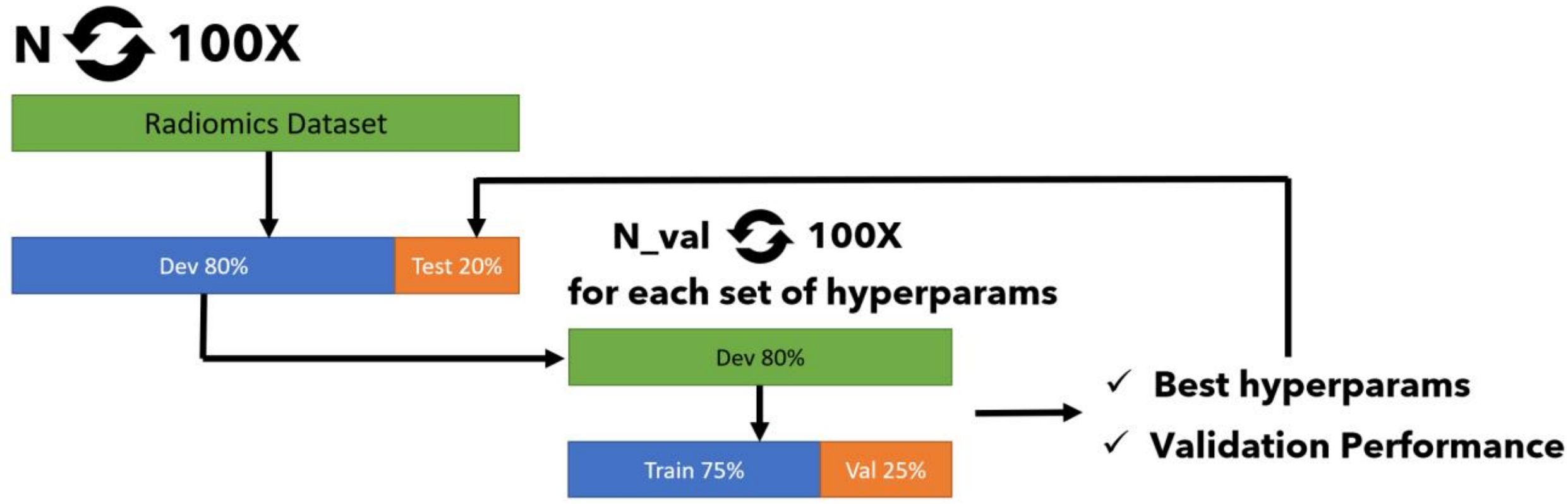

**Figure 3:** The repetitive classification approach

RF models are explainable and differentiative when applied to radiomics [33][34], and thus we chose them as our baseline classifiers. We defined a grid space for our classifier which is described in appendix B. For each combination of the hyperparameters in the grid space, we conduct N_val experiments (internal for loop). In this research, N_val was set to 100. In each experiment, the dev set is randomly split into training and validation (p_val%) cohorts. In this reseach, p_val is set to 25%. An instance of the classifier with the proposed hyperparameters is trained using the training set, and evaluated on the validation set based on the random split. While we use Area Under Receiver Operating Characteristic Curve (AUROC) as our evaluation metric, any other criterion, such as accuracy, may be utilized. Nonetheless, AUROC is a suitable metric for imbalanced datasets and medical research [35].

Once the N_val experiments are completed for the whole grid space, the best hyperparameter set is derived based on the highest average AUROC. To measure the validation performance, instances of models with the best hyperparameter set are trained and validated N_val times on random data splits with p_val% ratio. Average AUROC is considered to be the validation performance of the models. In the final step, an instance of the model with the best hyperparameters is trained on the entire dev set and evaluated on the test cohort. As it was mentioned, the whole process is repeated N times, and thus we have N test AUROCs as well as N validations AUROC.

Data management and other technical concerns such as deriving feature importance are discussed in appendices C and D, respectively. Table 1. includes the Open-radiomics technical protocol, and the settings for this study are provided appendix E.

**Table 1.** Open-radiomics technical research protocol

| | Technical consideration | Notes |
|---|---|---|
| 1 | Sources of variability for the research must be identified | |
| 2 | The sources of variability, whose effect will be measured, should be highlighted | |
| 3 | Authors should mention which tool was used for radiomics extraction | PyRadiomics is recommended |
| 4 | It must be confirmed all the required libraries were installed | In the case of PyRadiomics, the trimesh python library is required |
| 5 | All possible features should be extracted | Full set feature extraction is recommended. In the case of PyRadiomics, extractor.enableAllImageTypes() and extractor.enableAllFeatures() should be applied |
| 6 | It should be highlighted whether the study is 2D or 3D | In the case of 2D radiomics study on volumetric data, detailed process of deriving ROIs/VOIs should be included |
| 7 | The list of the extracted radiomics feature names must be provided in the supplementary materials | Providing the full list improved reproducibility of the research |
| 8 | A sample of the diagnostics features must be provided in the supplementary materials | Diagnostic features provide information about the environment that was used for feature extraction, improving repeatability and reproducibility of the research |
| 9 | The data split and model initialization must be repeated | With the small-size datasets, repeated train/validation/test splitting is recommended. If the number of repeats is low, e.g., 5-fold cross-validation, the training and validation indices should be provided |
| 10 | It must be highlighted if the pipeline included feature engineering | Any feature filtration, feature selection, or feature normalization technique should be learned from training or development sets and applied to the test set. Dataset-wise feature engineering may impact generalizability of the pipeline and thus lead to bias |
| 11 | The model, hyperparameter grid space, and evaluation metric(s) must be discussed in detailed | |
| 12 | It should be highlighted if the final model was trained on the development set or the training cohort | Retraining the model using the development set and the optimized hyperparameters is an effective approach for boosting the performance on small datasets. It can result in superior test performance compared to validation. |
| 13 | It is highly recommended the dataset (extracted radiomics features and the ground truth labels) be open-sourced | |
| 14 | The details of how the top-performing radiomics features were identified must be provided | |

## Data Records

All open-radiomics datasets are available on the project's website (https://openradiomics.org). Open-radiomics BraTS 2020 includes three compressed archives, encompassing the three binWidths (15, 25, and 35). Each archive contains 96 comma-separated values (CSV) files based on specific tumor subregions, image normalizations, and sequences. Due to higher volumes of data in BraTS 2023, 9 archives are curated. Thus, for each binWidth three archives are provided. Open-radiomics TCIA NSCLC is a single archive set. The CSV files for all three datasets are aligned with the Appendix C naming format. TCIA NSCLC preprocessed cohort is made available through Kaggle. Due to the dataset size limitations on the platform, the dataset is partitioned into two parts [22], [23].

# Results

## Classification Performance (brain tumors: HGG vs LGG)

Figure 4 a-d illustrates the effect of the four sources of variability on the AUROC performance of the classifiers, for validation and test cohorts. It should be highlighted that the highest AUROC might not be achieved with the combination of the best image normalization, binWidth, tumor subregion, and MRI sequence. Our approach is repetition-based and can be considered as a stochastic or random process, whose output is individual AUROCs. We achieve the highest average test AUROC performance ($0.956 \pm 0.033$) across the 100 experiments on T1CE, for NETnNCR subregion, with ZScore normalization, and with binWidth of 15. The second and the third top-performing datasets (mean test AUROCs of $0.955 \pm 0.030$ and $0.954 \pm 0.032$, respectively) had the same setting except for their image normalization, which was MinMax, and Gamma 1.5, respectively.

a)

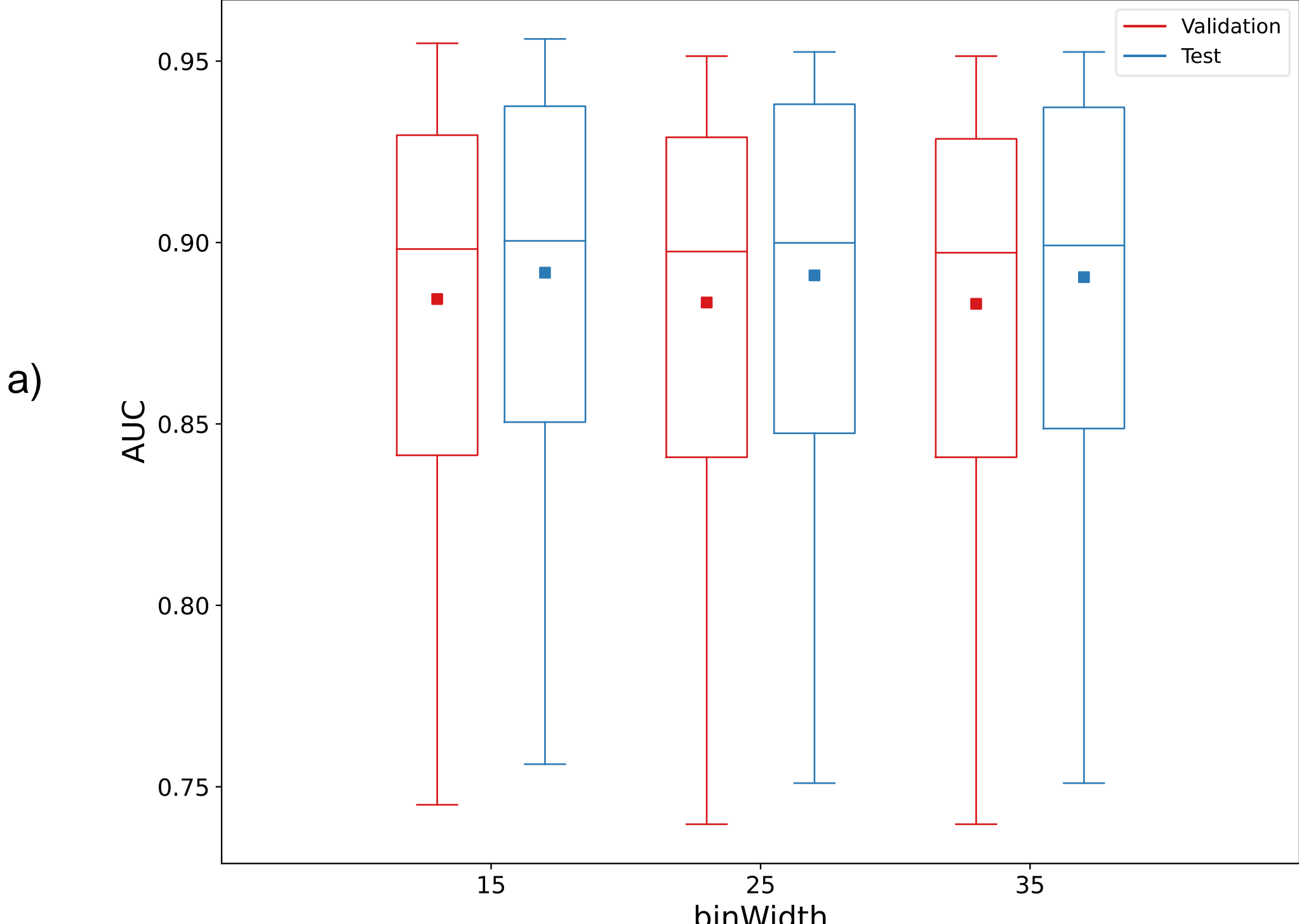

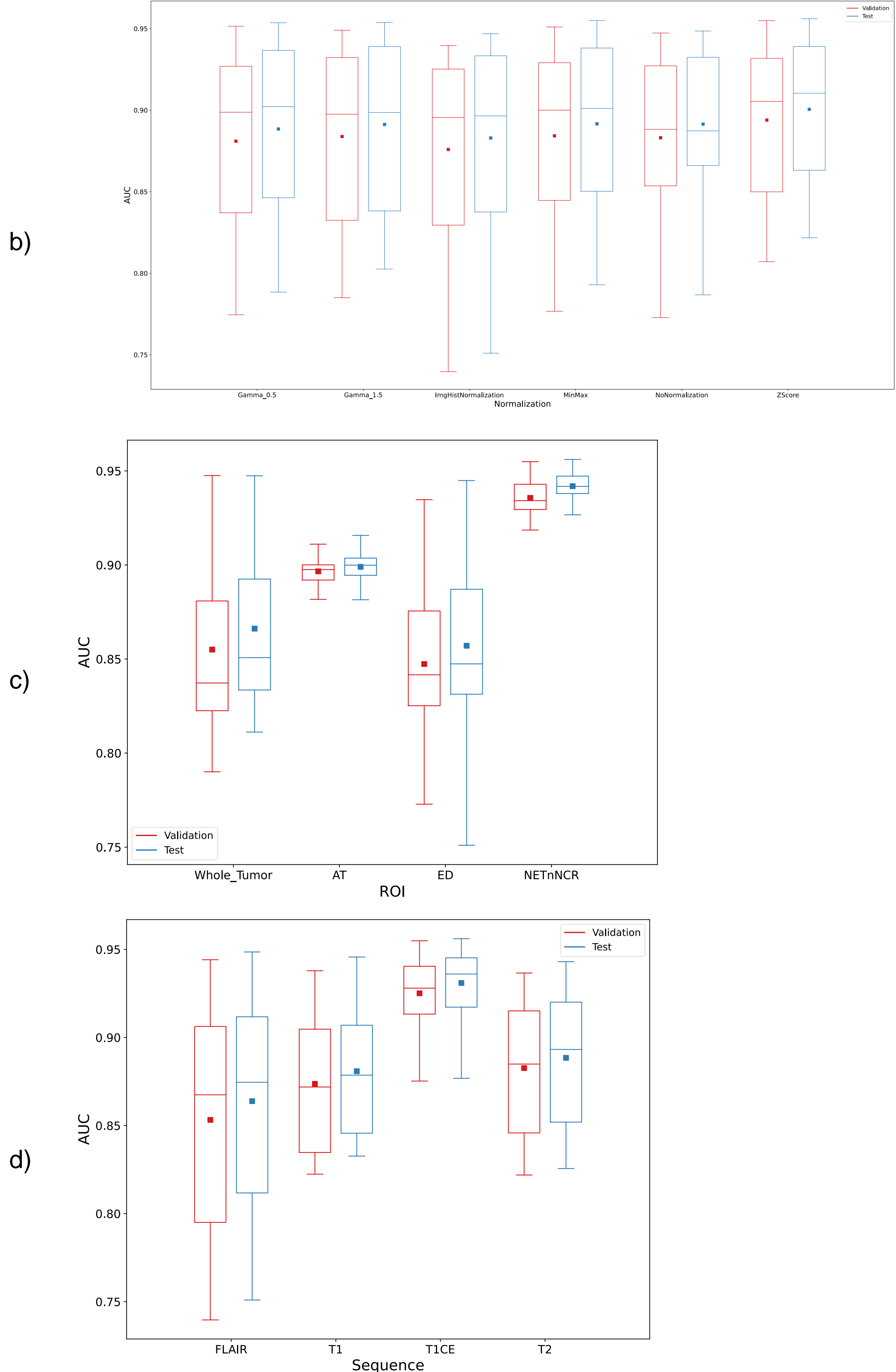

**Figure 4:** Effect of the studied factors on AUROC performance of the classifiers: a) binWidth b) image normalization c) VOI subregion d) MRI sequence

Figure 5 depicts how the top feature (lbp-3D-k_glszm_HighGrayLevelZoneEmphasis) differentiates the HGG and LGG examples on the top-performing dataset.

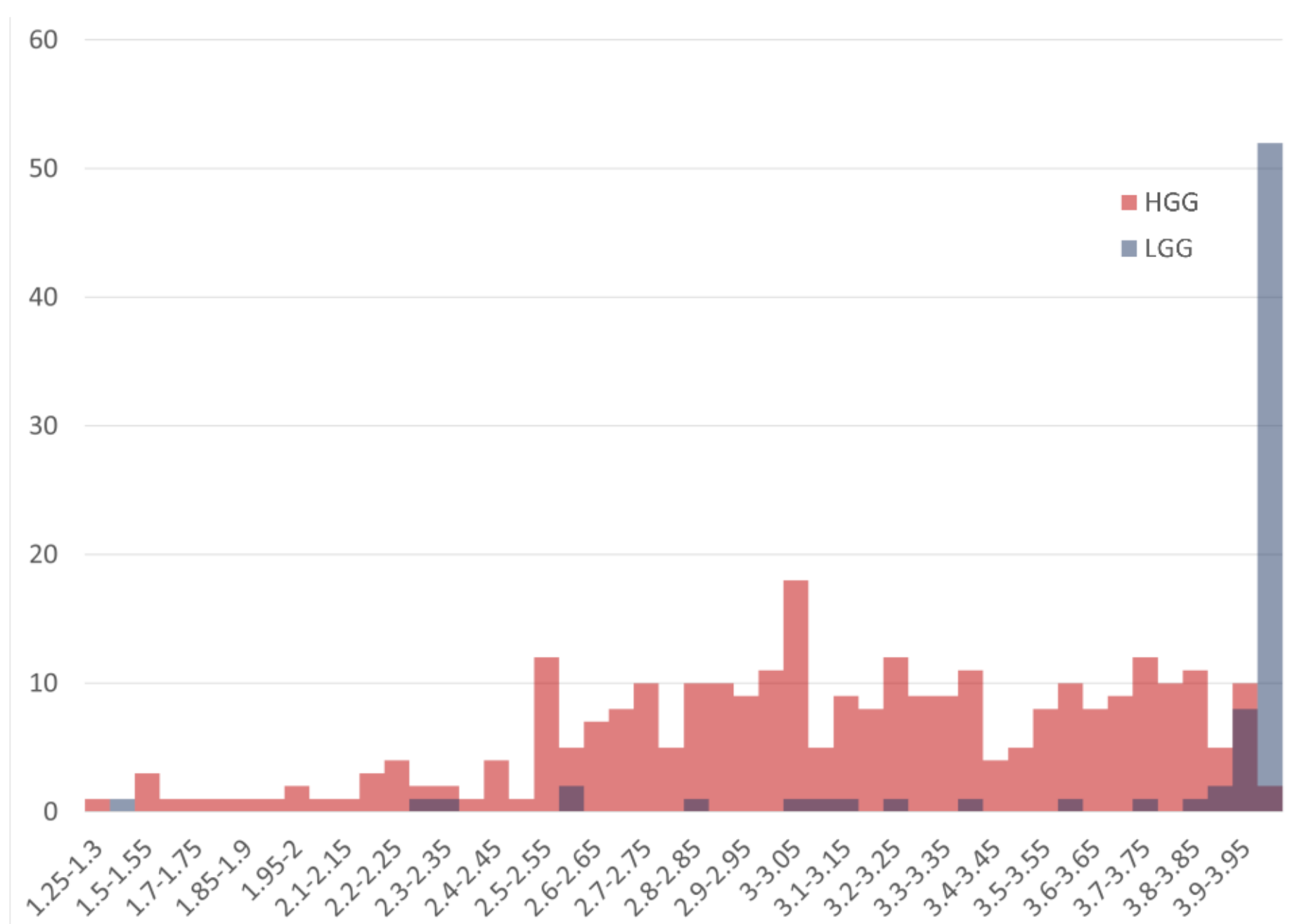

**Figure 5:** Histograms of the top feature on the top dataset: The horizontal axis represents bins of the feature values, and the vertical axis shows the number of VOIs with values in each bin

It is important to note that the boxplots in Figure 4 represent the range of average AUROCs. Hence, the maximum test performance of NETnNCR in Figure 4-c (AUROCs = 0.956) is itself an average of 100 experiments. Thus, there are multiple experiments in which we achieve very high AUROCs, which are close or equal to 1.00. Obviously, such high results are not reproducible. Along with the mean for each dataset, we also captured min, median, max, and first and third quartile of the AUROC performances. In 140 datasets out of 288, the highest achievable AUROC was above 0.99. In 28 cases, including T1CE, ZScore normalization, whole tumor classification, binWidth15, we reached to AUROC of 1.00. Nonetheless, our results prove that AUROC of 1.00 for the whole tumor classification is extremely irreproducible. Figure 6 illustrates the range of AUROC performance of the top-ten datasets.

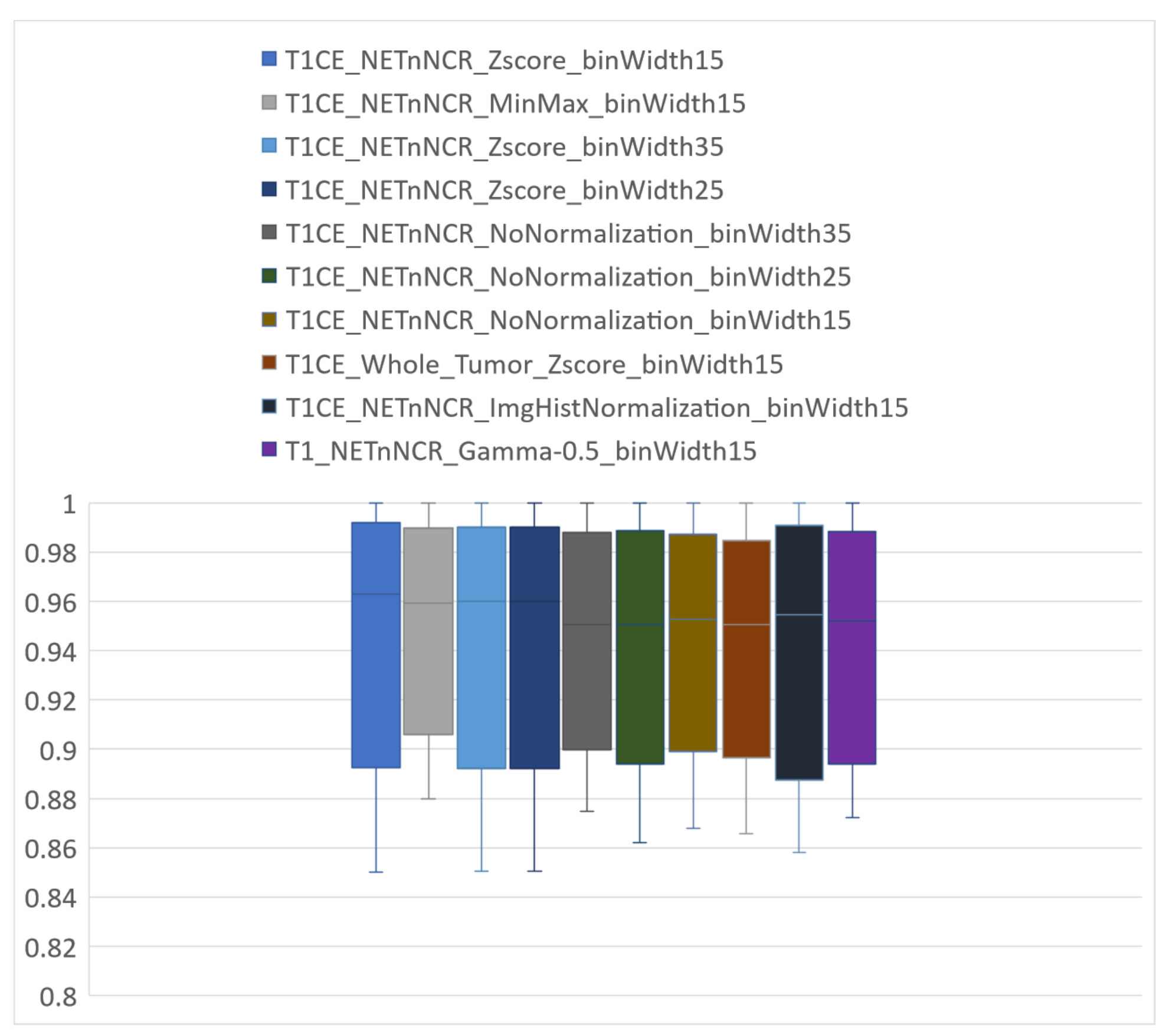

**Figure 6:** The 10 top-performing datasets

## Multisequence Classification Performance

We conducted a multisequence classification through combining the radiomics vectors of the four MRI sequences (T1, T1CE, T2, RLAIR). Figure 7 illustrates the effect of multisequence classification on improving the classification results. Multisequence classification improved the mean AUROC ($0.932 \pm 0.015$ compared with $0.891 \pm 0.048$ for single sequence, p-value < 0.001).

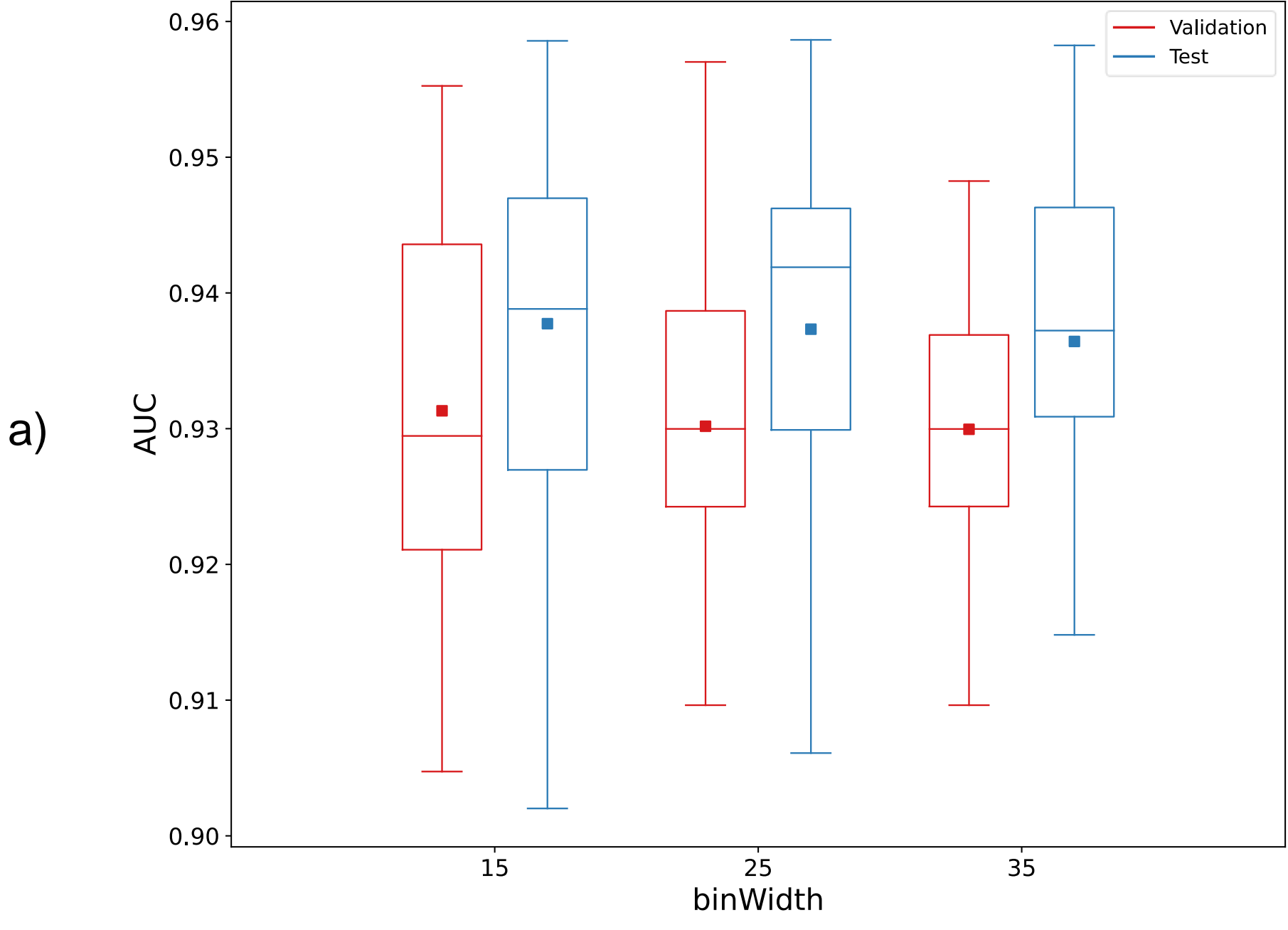

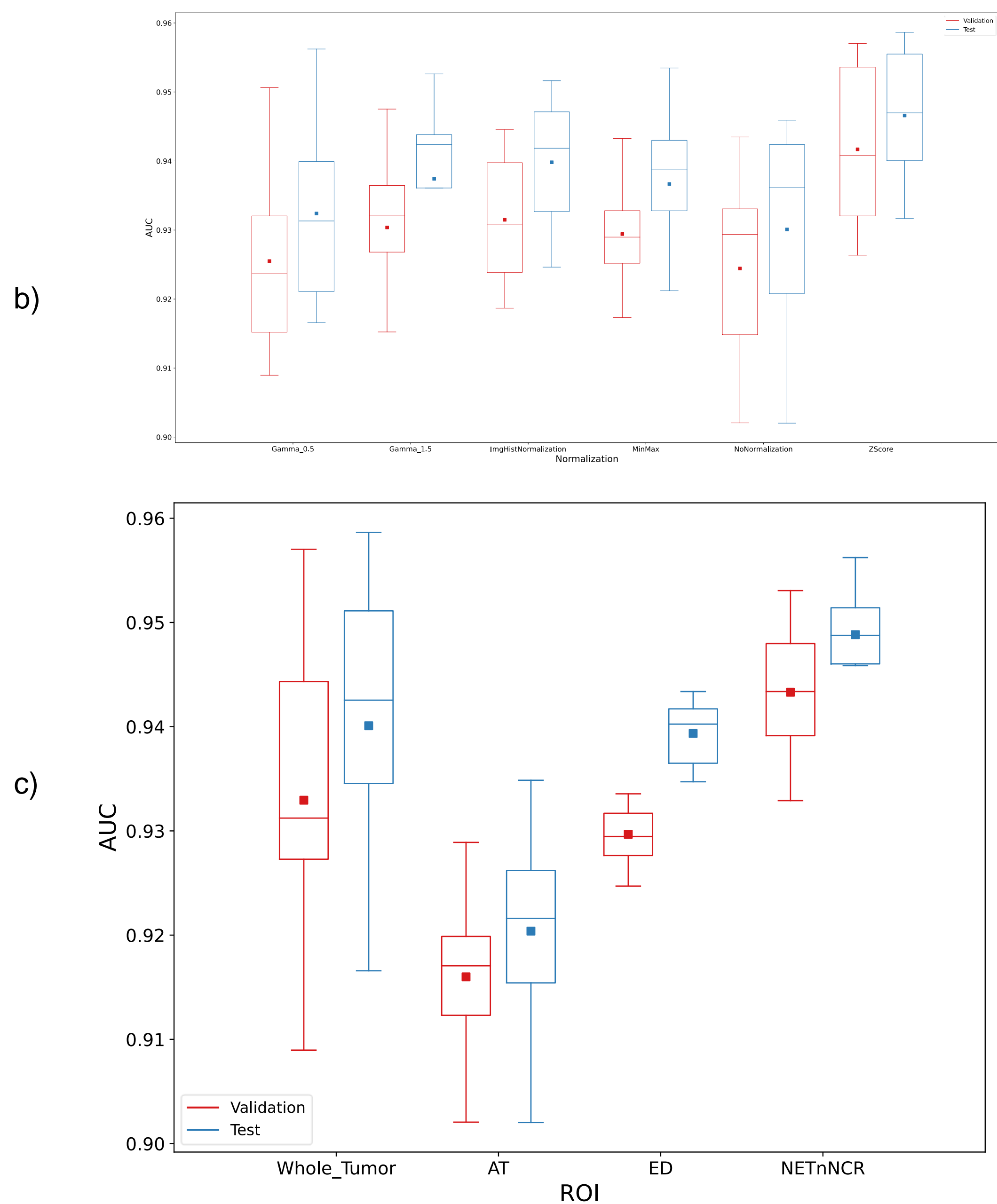

**Figure 7:** Effect of the studied factors on AUROC performance of the multisequence classifiers: a) binWidth b) image normalization c) VOI subregion

## Feature Extraction Failure

We evaluated the configurations where radiomics feature extraction failed entirely (see Appendix F). For BraTS 2020, a total of 696 out of 106,272 radiomic feature extractions were unsuccessful, with the optimal PyRadiomics binWidth appearing to be 25, the default setting. While the type of image normalization did not significantly influence failure rates (348 failed cases), the absence of normalization exacerbated the issue, leading to 445 failed cases. Feature extraction success was also dependent on tumor subregion characteristics, with larger and more convex subregions exhibiting higher success rates. Whole tumor classification had the fewest failures (48 cases), whereas the AT subregion resulted in the highest failure count (1952 cases). Although the differences among imaging sequences were marginal, T1 yielded the most reliable extractions, while T1CE exhibited the highest failure rate (535 vs. 556 cases, respectively).

### Top Ranking Feature

In our experiments, lbp-3D-k_glszm_HighGrayLevelZoneEmphasis radiomic feature appeared most frequently among the top features, and Figure 5 showed how this feature differentiated LGGs from HGGs. LBPs are operators that label pixels (or voxels, in the case of 3D VOIs) of images based on thresholding their neighbor points [24]. To form the lbp-3D-k, PyRadiomics extracts the spherical kurtosis image using the scipy.stats.kurtosis function [37], and applies the LBP operator to it. Kurtosis is defined as the fourth central moment times inverse of the square of the variance, and the kurtosis image corresponds to calculating kurtosis for every voxel. In radiomics, Gray Level Size Zone Matrix (GLSZM) is used for quantifying gray level zones in images. Gray level zones are defined as the number of connected voxels with similar gray-level intensities. High Gray Level Zone Emphasis (HGLZE), which is formulated as Eq. 1, represents a measurement of the distribution of the higher gray-level values. Higher HGLZE means the VOI contains a greater proportion of higher gray-level values and size zones.

$$HGLZE = \frac{\sum_{i=1}^{N_g} \sum_{j=1}^{N_s} \frac{P(i,j)}{i^2}}{N_z} \qquad \text{Eq. 1}$$

In Eq. 1., $N_g$ and $N_s$ correspond to the number of discrete intensity values, and the number of discrete zone sizes in the image, respectively. $N_z$ is the number of zones in the VOI, and $P_{(i,j)}$ is the size zone matrix.

Discussing a more comprehensive list of the top-performing features is out of the scope of this study. Nevertheless, we provide a list of the top ten features in supplemental material.

## Discussion

Our study highlights the impact of technical variability in radiomics-based machine learning pipelines, emphasizing the need for a standardized protocol to ensure reproducibility. By systematically examining imaging sequence selection, binWidth values, image normalization techniques, and tumor subregion choices, we provide a comprehensive assessment of their effects on classification performance.

Overall, binWidth does not make a tangible difference in classification performance, with average test AUROCs of 0.892, 0.891, and 0.890 for binWidth values of 15, 25, and 35, respectively. Among the image normalization techniques examined, Z-score normalization slightly improved the average model performance (AUROC = 0.901) while also reducing the variance of test AUROCs (average standard deviation (SD) = 0.042), making it a preferable choice for radiomics-based ML pipelines. Our results suggest that the true potential of radiomics may not be fully exploited for whole tumor classification in BraTS 2020, as specific tumor subregions yielded better results. In particular, subregion selection significantly influenced model performance, with NETnNCR producing the highest AUROCs. This finding supports prior research indicating that tumor subregion delineation is critical in radiomics-based classification tasks.

Additionally, we observed that T1CE consistently outperformed other imaging sequences, further underscoring its importance in radiomics pipelines. Among the four MRI sequences, T1CE achieved the highest performance (AUROC = 0.931), whereas FLAIR had the lowest performance (AUROC = 0.864). The classification models for NETnNCR subregions of HGG and LGG tumors demonstrated high accuracy and reproducibility, with an average test AUROC of 0.942. In contrast, models for the AT subregion had an average test AUROC of 0.899 but exhibited low variance (SD = 0.010), indicating stable results. In the case of ED, we observed the highest AUROC variance (SD = 0.055), suggesting that its classification results were less reliable. Finally, the multisequence classification demonstrated that combining multiple MRI sequences led to a performance boost, reinforcing the benefit of leveraging multi-sequence imaging for radiomics-based machine learning models.

Except for the shape and first-order features, explaining the details and physical meaning of radiomics features is not straightforward. Radiomics includes multiple groups of features, which are described in the Image Biomarker Standardization Initiative (IBSI) [36]. One of the most frequently occurring top features in our experiments was lbp-3D-k_glszm_HighGrayLevelZoneEmphasis, highlighting the importance of LBP and GLSZM features in brain tumor classification. While radiomics features provide a degree of explainability not typically found in deep learning models, the clinical relevance of these features requires further validation.

The analysis of stochastic variability across multiple repetitions (N = 100) confirmed that high AUROC values (approaching 1.00) in radiomics studies are often irreproducible. This underscores the importance of repeated evaluations to differentiate between stable and overfitted results. By incorporating repetition-based analysis, we reduce the likelihood of misleading findings that may result from a single train-test split.

Unlike prior work, our study systematically investigates multiple technical factors influencing reproducibility in radiomics classification, providing a rigorous baseline for future research. The findings also align with prior research on inter-reader variability and scanner-specific biases in radiomics, further supporting the necessity of standardized protocols.

Although our study provides valuable insights into technical variability, several limitations must be acknowledged. While PyRadiomics is a widely used tool, feature extraction discrepancies across different radiomics software remain an open challenge. Future work should explore additional feature extraction frameworks to validate our protocol across multiple platforms. Furthermore, incorporating prospective datasets with multi-center variability will help assess real-world reproducibility beyond retrospective data. Finally, while we prioritize reproducibility over model optimization, integrating explainable AI techniques could enhance feature interpretability and facilitate clinical adoption.

## Conclusions

In this research, we proposed a research protocol for radiomics-based ML pipelines to improve reproducibility. We extracted and open-sourced a large series of tabular radiomics datasets based on the BraTS 2020 dataset that enables multiple opportunities for radiomics research. We established a reproducible baseline for the open-radiomics datasets, and studied the effect of PyRadiomics binWidth, image normalization, VOI subregion, and the MRI sequence as four sources of variability on the performance of the radiomics pipeline.

Our experiments demonstrated the default binWidth increases the chance of a successful VOI feature extraction, while it has no effect on the generalizability of the model. For HGG versus LGG classification, NETnNCR VOI subregion is associated with the highest-performing models, and AT ranks second. However, ED and whole tumor classifications struggle to achieve comparable performance. We found T1CE and FLAIR to be the best and worst sequences, respectively. While these results may be specific to BraTS dataset, the protocol can be followed to generate reliable and reproducible radiomics results for any given dataset.

## Acknowledgements

This research has been made possible with the financial support of the Canadian Institutes of Health Research (CIHR) (Funding Reference Number: 184015)

# Appendix A. How 2D features are extracted from VOIs

When full-set feature extraction for a VOI (3D) is enabled, 2D LBP features are extracted but 2D shape features are skipped by PyRadiomics. With the default settings, 2D LBPs are calculated for each slice of the VOI in the axial direction and stacked. The following code snippet from PyRadiomics shows how 2D LBP features are calculated for 3D VOIs (https://github.com/AIM-Harvard/pyradiomics/blob/master/radiomics/imageoperations.py).

```
lbp_radius = kwargs.get('lbp2DRadius', 1)
lbp_samples = kwargs.get('lbp2DSamples', 8)
lbp_method = kwargs.get('lbp2DMethod', 'uniform')

im_arr = sitk.GetArrayFromImage(inputImage)

Nd = inputImage.GetDimension()
if Nd == 3:
    # Warn the user if features are extracted in 3D, as this function calculates LBP in 2D
    if not kwargs.get('force2D', False):
      logger.warning('Calculating Local Binary Pattern in 2D, but extracting features in 3D.
Use with caution!')
    lbp_axis = kwargs.get('force2Ddimension', 0)

    im_arr = im_arr.swapaxes(0, lbp_axis)
    for idx in range(im_arr.shape[0]):
      im_arr[idx, ...] = local_binary_pattern(im_arr[idx, ...],
                                              P=lbp_samples,
                                              R=lbp_radius,
                                              method=lbp_method)
    im_arr = im_arr.swapaxes(0, lbp_axis)
```

# Appendix B. Grid space of the RF models

**Table 2.** Grid space of the RF models

| Hyperparameter | Grid Space |
|---|---|
| n_estimators | 50, 100, 200 |
| max_features | 'auto', 'sqrt' |
| max_depth | None, 5, 10 |

# Appendix C. Data management

Data management is an essential part of radiomics studies. To avoid data fragmentation and possible mistakes, we suggest saving the features, ground truth labels, clinical variables, and any other information in a single csv file. Each row of the csv file (except the header row) should belong to a unique ROI/VOI. For the studies where a patient might have multiple ROIs/VOIs, creating unique ROI/VOI IDs and appending them as the first column to the dataset is preferred. The naming of the radiomics datasets (csv files) is important. All names start with “Radiomics”, and the format is always csv. We use ‘_’ as the separator to include type of the normalization, and sequence in the naming. If needed, other pieces of information can be included. Obviously, no underscore should be used within the parts. Two examples of naming would be Radiomics_Gamma-0.5_FLAIR.csv, and Radiomics_NoNormalization_ED_T1.csv.
In this research, we study the effect of image normalization, imaging sequence, tumor subregion, and PyRadiomics settings on an adult brain tumor classification pipeline.

As it was mentioned, we have 6 image normalization methods (i.e., NoNormalization, Gamma0.5, Gamma1.5, Histogram, ZScore, and MinMax), 4 imaging sequences (T1, T1CE, T2, FLAIR), 4 tumor subregions (i.e., whole tumor, AT, ED, and NETnNCR), 3 different binWidth values (i.e., 15, 25, and 35). This creates 288 sets of tabular datasets for the radiomic features (~2.9 GB of data).

## Appendix D. Other technical concerns

Although no randomness is involved in the feature extraction process, seeding is recommended in all the codes. This improves reproducibility of the results if random IDs are assigned to ROIs/VOIs, or further analysis such as dimensionality reduction and data visualization are incorporated into the scripts. We encourage extracting the full set of radiomics features, which is not enabled by default ( extractor.enableAllImageTypes() and extractor.enableAllFeatures() will result in a full set feature extraction).
Radiomics-based ML pipelines are more explainable compared with DL, which is a result of the transparent definitions of the radiomic features. Radiomics studies usually are concluded by highlighting the most important features, and we encourage this approach. However, not every algorithm is explainable. As an example, once an NN classifies a radiomics example, determining the influential features is perplexing. RF is an explainable model, and thus feature importances of an RF classifier can be derived. We capture the most important feature of each N experiments, for each dataset. Hence, we will have a list of $288 \times N$ top features, and the most frequent element of the list will represent the number one radiomics feature for BraTS 2020 tumor type classification. The last technical point is the choice of N=100 and N_val=100, which is made based on the practice of the Central Limit Theorem (CLT). This number of repetitions eliminates the need for k-fold cross-validation. Hence, we suggest setting N and N_vals above 30, where computational costs allow, and switching to k-fold cross-validation, otherwise.

## Appendix E. Open-radiomics settings for the current research

**Table 3.** Open-radiomics settings for the current research

| | Technical consideration | Setting for the current research |
|---|---|---|
| 1 | Sources of variability for the research must be identified | Segmentation (intra- and inter-reader variability), Imaging scanner vendor, imaging protocol, Imaging sequence, binWidth, image normalization, tumor subregion |
| 2 | The sources of variability, whose effect will be measured, should be highlighted | Imaging sequence, binWidth, image normalization, tumor subregion |
| 3 | Authors should mention which tool was used for radiomics extraction | PyRadiomics |
| 4 | It must be confirmed all the required libraries were installed | check |
| 5 | All possible features should be extracted | check |

| 6 | It should be highlighted whether the study was 2D or 3D | 3D |
|---|---|---|
| 7 | The list of the extracted radiomics feature names must be provided in the supplementary materials | check |
| 8 | A sample of the diagnostics features must be provided in the supplementary materials | check |
| 9 | The data split and model initialization must be repeated | yes, we employed repeated train/validation/test and set different random states for the models with each split |
| 10 | It must be highlighted if the pipeline included feature engineering | yes, the feature engineering was learned from the development set |
| 11 | The model, hyperparameter grid space, and evaluation metric(s) must be discussed in detailed | check |
| 12 | It should be highlighted if the final model was trained on the development set or the training cohort | The final models were trained on the developments sets |
| 13 | It is highly recommended the dataset (extracted radiomics features and the ground truth labels) be open-sourced | yes |
| 14 | The details of how the top-performing radiomics features were identified must be provided | We selected the top features based on RF feature importance scores |

## Appendix F. Feature Extraction Failure

We define the failure as a fatal error or a timeout produced by the PyRadiomics library during the feature extraction. On a system with an AMD Ryzen threadripper pro 3955wx, 128 GB of RAM, 4 TB of M2 SSD, running Ubuntu 20.04.4 LTS, we set the timeout threshold at 120 seconds which created a safe margin because common ROI/VOI feature extraction time was of the order of seconds. Corresponding figures are provided in supplemental material. It should be highlighted that identifying the reason for feature extraction failure is out of the scope of this research.

## Supplemental Material

https://openradiomics.org?download=1209&tmstv=1718853413